\documentclass[aps,prd,twocolumn,10pt,groupedaddress]{revtex4-1}
\usepackage{amssymb}
\usepackage{graphicx}
\usepackage{amsmath}
\usepackage{hyperref}
\usepackage{subfigure}
\usepackage{float}
\usepackage{color}
\usepackage{ulem}
\usepackage[utf8]{inputenc}

\begin{document}

\title{Higgs chameleon}
\author{Rong-Gen Cai}
\email{cairg@itp.ac.cn}
\affiliation{CAS Key Laboratory of Theoretical Physics, Institute of Theoretical Physics, Chinese Academy of Sciences, Beijing 100190, China}
\affiliation{School of Fundamental Physics and Mathematical Sciences, Hangzhou Institute for Advanced Study, University of Chinese Academy of Sciences, Hangzhou 310024, China}
\author{Shao-Jiang Wang}
\email{schwang@cosmos.phy.tufts.edu (corresponding author)}
\affiliation{Tufts Institute of Cosmology, Department of Physics and Astronomy, Tufts University, 574 Boston Avenue, Medford, Massachusetts 02155, USA}

\begin{abstract}
The existing constraints from particle colliders reveal a suspicious but  nonlethal metastability for our current electroweak vacuum of Higgs potential in the standard model of particle physics, which is, however, disfavored in the early Universe if the inflationary Hubble scale is larger than the instability scale when Higgs quartic self-coupling runs into negative value. Alternative to previous trials of acquiring a positive effective mass-squared from Higgs quadratic couplings to Ricci scalar or inflaton field, we propose a third approach to stabilize the Higgs potential in the early Universe by regarding Higgs as chameleon coupled to inflaton alone without conflicting to the present constraints on either Higgs or chameleon.
\end{abstract}
\maketitle

\section{Introduction}\label{sec:int}

The state-of-art measurements \cite{Tanabashi:2018oca} on Higgs mass $M_h=125.10\pm0.14\, \mathrm{GeV}$ and top quark mass $M_t=172.9\pm0.4\,\mathrm{GeV}$ continue to reenforce the longstanding conspiracy of Higgs near-criticality \cite{EliasMiro:2011aa,Bezrukov:2012sa,Degrassi:2012ry,Alekhin:2012py,Buttazzo:2013uya} (see also \cite{Espinosa:2018mfn,Markkanen:2018pdo} for recent reviews and references therein). The running of Higgs quartic self-coupling starts becoming negative around the dubbed instability scale $\Lambda_I=9.92\times10^9\,\mathrm{GeV}$ \cite{Bednyakov:2015sca} (see also \cite{DiLuzio:2014bua,Andreassen:2014eha,Andreassen:2014gha,Espinosa:2016nld} for its gauge dependence), where the Higgs potential develops a shallow barrier unstable against quantum fluctuations of order $H_\mathrm{inf}/(2\pi)$ during inflation if the inflationary Hubble scale $H_\mathrm{inf}$ is larger than $\Lambda_I$. Therefore, the survival of our current electroweak (EW) vacuum throughout a high scale inflation seems highly unnatural and undesirable, even though we are temporarily safe in the EW vacuum for a lifetime of order $10^{161}\,\mathrm{yrs}$ \cite{Andreassen:2017rzq} against Coleman-de Luccia (CdL) instanton with decay rate estimated around $10^{-554}\,\mathrm{Gyr}^{-1}\mathrm{Gpc}^{-3}$ \cite{Chigusa:2017dux,Chigusa:2018uuj} (see also \cite{Figueroa:2015rqa} for lattice simulation result and \cite{Rose:2015lna} for most recent results with thermal corrections). This is known as Higgs metastability, a special case of Higgs near-criticality, since the running of Higgs quartic self-coupling could otherwise be fairly stable all the way to Planck scale within the current uncertainties mainly from top quark mass and strong coupling.

The attitude toward Higgs near-criticality could be either desirable or deniable. In the former case, the Higgs near-criticality could be the plausible smoking gun for the possible ultraviolet completion of the standard model (SM) of particle physics, for example, asymptotic safe gravity \cite{Shaposhnikov:2009pv}\footnote{It is worth mention that the scenario of asymptotic safe gravity predicts Higgs mass around 126 GeV. Similar prediction was also achieved in \cite{Liu:2012qua} from high scale supersymmetry.}, metastable Higgs inflation \cite{Bezrukov:2014ipa}, dynamical criticality \cite{Khoury:2019ajl}, to name just a few. In the latter case, the Higgs near-criticality could also be a mirage for our ignorance of new physics, for example, the Planckian physics with higher-order Higgs self-interactions \cite{Branchina:2013jra,Branchina:2014usa,Branchina:2014rva,Lalak:2014qua,Branchina:2018xdh,Branchina:2019tyy} or Planck-suppressed derivative operators \cite{Fumagalli:2019ohr}, and the extra contributions to Higgs effective mass-squared during inflation from the quadratic coupling to inflaton field \cite{Kobakhidze:2013tn,Fairbairn:2014zia,Kamada:2014ufa} (see also \cite{Hertzberg:2019prp}) or the nonminimal coupling to Ricci scalar \cite{Espinosa:2007qp,Herranen:2014cua,Czerwinska:2016fky,Markkanen:2017dlc}. The corresponding postinflationary investigations \cite{Herranen:2015ima,Ema:2016kpf,Kohri:2016wof,Kohri:2016qqv,Enqvist:2016mqj,Postma:2017hbk,Ema:2017loe,Kohri:2017iyl,Ema:2017rkk,Figueroa:2017slm}  are also crucial for the eventual fate determination \cite{Hook:2014uia,Espinosa:2015qea,East:2016anr}. Although the gravitational corrections to Higgs decay from EW vacuum are negligible \cite{Isidori:2007vm,Branchina:2016bws,Rajantie:2016hkj,Salvio:2016mvj,Joti:2017fwe,Rajantie:2017ajw,Espinosa:2020qtq}, the catalyzed vacuum decay by black holes \cite{Gregory:2013hja,Burda:2015isa,Burda:2015yfa,Burda:2016mou,Gorbunov:2017fhq,Canko:2017ebb,Kohri:2017ybt,Gregory:2018bdt,Oshita:2019jan} (see \cite{Oshita:2016oqn,Mukaida:2017bgd} for its thermal interpretation and  \cite{Gregory:2020cvy} for its thermal extension) or other compact objects \cite{Oshita:2018ptr}, braneworld \cite{Cuspinera:2018woe,Cuspinera:2019jwt,Mack:2018fny}, cosmic string \cite{Koga:2019mee,Koga:2019yzj,Firouzjahi:2020hfq} and naked singularity \cite{Oshita:2020ksb} should be of special concern. Similar consideration of excited initial states at false vacuum \cite{Darme:2017wvu} could also affect the decay rate, even possibly in real-time \cite{Braden:2018tky,Hertzberg:2019wgx,Blanco-Pillado:2019xny,Darme:2019ubo,Wang:2019hjx,Huang:2020bzb}.

Inspired by the chameleon mechanism \cite{Khoury:2003aq,Khoury:2003rn,Wang:2012kj,Upadhye:2012vh,Khoury:2013yya} by coupling the chameleon to ambient matter where the effective potential of chameleon becomes heavier in the denser environment, we propose in Sec. \ref{sec:HiggsChameleon} to stabilize the Higgs field in the early Universe by recognizing Higgs as chameleon coupled to inflaton after we first generalize the chameleon coupling for arbitrary background in Sec. \ref{sec:GeneralChameleon}. The idea is simple enough but has never been explored before \footnote{The chameleon was proposed as an effective screening mechanism for modified gravity, and has been widely used to account for dark energy or even dark matter (see, for example,  \cite{Katsuragawa:2016yir,Katsuragawa:2018wbe,Chen:2019kcu}).}, which is also free from all the current constraints on Higgs from particle colliders and on chameleon from local gravity experiments if we restrict ourselves to couple Higgs chameleon to inflaton alone.

\section{Higgs as chameleon}\label{sec:GeneralChameleon}

Choosing the scalar field $h$ as the chameleon field introduces extra interactions between $h$ and other matter fields $\psi_i$ with action in the Einstein frame of form
\begin{align}
S=&\int\mathrm{d}^4x\sqrt{-g}\left(\frac{M_{\mathrm{Pl}}^2}{2}R-\frac{1}{2}(\partial h)^2-V(h)\right)\nonumber\\
&+\sum\limits_iS^{(i)}_\mathrm{m}\left[\Omega_i^2(h)g_{\mu\nu},\psi_i\right],
\end{align}
where the reduced Planck mass $M_\mathrm{Pl}^2=(8\pi G)^{-1}$ and the chameleon couplings to the metric $g_{\mu\nu}$ induce new metrics $\tilde{g}^{(i)}_{\mu\nu}=\Omega_i^2(h)g_{\mu\nu}$ for each fields $\psi_i$ that are assumed to be independent for simplicity. The corresponding action variation (the variations $\delta\psi_i$ are not shown here) reads
\begin{align}
\delta S&=\int\mathrm{d}^4x\sqrt{-g}\left(\frac{M_\mathrm{Pl}^2}{2}G_{\mu\nu}-\frac12T_{\mu\nu}^{(h)}\right)\delta g^{\mu\nu}\label{eq:dSEH}\\
&+\int\mathrm{d}^4x\sqrt{-g}\left(\nabla^2h-V'(h)\right)\delta h\label{eq:dSh}\\
&+\sum\limits_i\int\mathrm{d}^4x\sqrt{-\tilde{g}_{(i)}}\left(-\frac12\tilde{T}_{\mu\nu}^{(i)}\right)\delta\tilde{g}^{\mu\nu}_{(i)}\label{eq:dSm}
\end{align}
with the Einstein tensor $G_{\mu\nu}\equiv R_{\mu\nu}-\frac12g_{\mu\nu}R$ and the energy-momentum tensors defined by
\begin{align}
T_{\mu\nu}^{(h)}&=\frac{-2}{\sqrt{-g}}\frac{\delta S_h}{\delta g^{\mu\nu}}=\frac{-2}{\sqrt{-g}}\frac{\partial(\sqrt{-g}\mathcal{L}_h)}{\partial g^{\mu\nu}}\nonumber\\
&=\nabla_\mu h\nabla_\nu h+g_{\mu\nu}\left(-\frac{1}{2}(\partial h)^2-V(h)\right),\\
\tilde{T}_{\mu\nu}^{(i)}&=\frac{-2}{\sqrt{-\tilde{g}_{(i)}}}\frac{\partial}{\partial\tilde{g}^{\mu\nu}_{(i)}}\left(\sqrt{-\tilde{g}_{(i)}}\mathcal{L}_\mathrm{m}^{(i)}[\tilde{g}_{\mu\nu}^{(i)},\psi_i]\right),
\end{align}
where the last contribution \eqref{eq:dSm} could be rewritten with respect to the Einstein-frame metric as
\begin{align}
&\sum\limits_i\int\mathrm{d}^4x\sqrt{-g}\,\Omega_i^4\left(-\frac12\tilde{T}_{\mu\nu}^{(i)}\right)\left(\Omega_i^{-2}\delta g^{\mu\nu}-\frac{2\Omega'_i(h)}{\Omega_i^3}g^{\mu\nu}\delta h\right)\nonumber\\
&=\sum\limits_i\int\mathrm{d}^4x\sqrt{-g}\left(-\frac12\tilde{T}_{\mu\nu}^{(i)}\Omega_i^2\delta g^{\mu\nu}+\Omega'_i(h)\Omega_i^3\tilde{T}_i\delta h\right)\label{eq:dSm1}
\end{align}
with trace $\tilde{T}_i\equiv\tilde{T}_{\mu\nu}^{(i)}\tilde{g}^{\mu\nu}_{(i)}$. On the other hand, $\delta S_\mathrm{m}$ could also be expressed in terms of chain rule as
\begin{align}
&\sum\limits_i\int\mathrm{d}^4x\left(\frac{\delta S_\mathrm{m}^{(i)}}{\delta g^{\mu\nu}}\delta g^{\mu\nu}+\frac{\delta S_\mathrm{m}^{(i)}}{\delta\tilde{g}^{\mu\nu}_{(i)}}\frac{\delta\tilde{g}^{\mu\nu}_{(i)}}{\delta h}\delta h\right)\nonumber\\
&=\sum\limits_i\int\mathrm{d}^4x\sqrt{-g}\left(-\frac12T_{\mu\nu}^{(i)}\delta g^{\mu\nu}+\Omega'_i(h)\Omega_i^3\tilde{T}_i\delta h\right),\label{eq:dSm2}
\end{align}
which, after compared with \eqref{eq:dSm1}, leads to identification
\begin{align}
\tilde{T}_{\mu\nu}^{(i)}\Omega_i^2=T_{\mu\nu}^{(i)}\equiv\frac{-2}{\sqrt{-g}}\frac{\delta S_\mathrm{m}^{(i)}}{\delta g^{\mu\nu}}.
\end{align}
Thus $\tilde{T}_{(i)}^{\mu\nu}\Omega_i^6=T_{(i)}^{\mu\nu}$, $\tilde{T}^\mu_{(i)\nu}\Omega_i^4=T^\mu_{(i)\nu}$ and $\tilde{T}_i\Omega_i^4=T_i$. 

The energy-momentum tensor $\tilde{T}^{\mu\nu}_{(i)}$ is conserved by $\tilde{\nabla}_\mu^{(i)}\tilde{T}^{\mu\nu}_{(i)}=0$ in Jordan frame where $\psi_i$ is minimally coupled to the Jordan-frame metric $\tilde{g}_{\mu\nu}^{(i)}$. However, the energy-momentum tensor is not conserved as $\nabla_\mu T_{(i)}^{\mu\nu}=0$ in Einstein frame. In fact, note that $\tilde{\Gamma}^{\rho(i)}_{\mu\nu}=\Gamma^\rho_{\mu\nu}+C^{\rho(i)}_{\mu\nu}$ with $C^{\rho(i)}_{\mu\nu}=\Omega_i^{-1}(\delta^\rho_\mu\nabla_\nu\Omega_i+\delta^\rho_\nu\nabla_\mu\Omega_i-g_{\mu\nu}g^{\rho\lambda}\nabla_\lambda\Omega_i)$, we have $\tilde{\nabla}_\mu^{(i)}\tilde{T}^{\mu\nu}_{(i)}=\Omega_i^{-6}\nabla_\mu T^{\mu\nu}_{(i)}-T_i\Omega_i^{-7}\nabla^\nu\Omega_i=0$, namely,
\begin{align}\label{eq:conservation}
\nabla_\mu T^\mu_{(i)\nu}=T_i\Omega_i^{-1}\nabla_\nu\Omega_i,
\end{align}
For a perfect fluid ansatz for $T^\mu_{(i)\nu}=\mathrm{diag}(-\rho_i,p_i,p_i,p_i)$ with equation-of-state (EoS) parameter $w_i$ defined by $p_i=w_i\rho_i$, the $\nu=0$ component of \eqref{eq:conservation} reads $\nabla_t\rho_i=(1-3w_i)\rho_i\nabla_t\ln\Omega_i$,
which could be rearranged into
\begin{align}
\nabla_t\left(\Omega_i^{3w_i-1}\rho_i\right)=0
\end{align}
if EoS parameter $w_i$ is treated as a constant. This defines a covariantly conserved density in the Einstein frame by 
\begin{align}
\hat{\rho}_i=\Omega_i^{3w_i-1}\rho_i=\Omega_i^{3w_i+3}\tilde{\rho}_i,
\end{align}
which is also $h$-independent from $0=\nabla_t\hat{\rho}_i=\hat{\rho}'_h(h)\nabla_th$. Now requiring vanishing variation for the sum of \eqref{eq:dSEH}, \eqref{eq:dSh} and \eqref{eq:dSm1} gives rise to the equation-of-motions (EoMs) for the metric field $g_{\mu\nu}$ and scalar field $h$ as
\begin{align}
G_{\mu\nu}&=8\pi G\left(T_{\mu\nu}^{(h)}+\sum\limits_iT_{\mu\nu}^{(i)}\right),\label{eq:Einstein}\\
\nabla^2h&=V'(h)-\sum\limits_i\Omega'_i(h)\Omega_i^3(h)\tilde{T}_i,\label{eq:HiggsEoM}
\end{align}
where the scalar EoM \eqref{eq:HiggsEoM} could be rewritten as $\nabla^2h=V'_\mathrm{eff}(h)$ with respect to an effective potential $V_\mathrm{eff}(h)=V(h)+\sum\limits_iU_i(h)$ with $U_i(h)$ of form
\begin{align}\label{eq:Uh}
U_i(h)=\Omega_i^{1-3w_i}(h)\hat{\rho}_i=\begin{cases}
\Omega_\mathrm{vac}^4\hat{\rho}_\mathrm{vac} &,\, i=\text{vacuum energy}\\
\hat{\rho}_r &,\, i=\text{radiation} ;\\
\Omega_m\hat{\rho}_m &,\, i=\text{matter}.
\end{cases}
\end{align}
Note that for radiation domination, $\hat{\rho}$ is covariantly constant in time and hence $h$-independent. Hereafter, we will choose the scalar  field $h$ as Higgs field specifically.

\section{Higgs chameleon in the early Universe}\label{sec:HiggsChameleon}

For the sake of simplicity, the Higgs field is assumed to have no chameleon coupling to all the other fields except inflaton field, then the Higgs effective potential $V_\mathrm{eff}$ only receives its contribution of $U_i$  from inflaton field alone as
\begin{align}\label{eq:HiggsChameleon}
V_\mathrm{eff}(h)=V(h)+\hat{\rho}_\phi\Omega_\phi^{1-3w_\phi}(h).
\end{align}
The SM Higgs potential at zero temperature with higher loop-order quantum corrections could be approximated as \cite{Espinosa:2015qea}
\begin{align}
V(h)=V_0(h)\approx-b\log\left(\frac{h^2}{h_c^2\sqrt{e}}\right)\frac{h^4}{4},
\end{align}
where the Higgs quartic coupling turns negative at a critical value $h_c\simeq5\times10^{10}$ GeV and $b\approx0.16/(4\pi)^2$. To save Higgs from the instability developed around $h_c$, there are infinitely many choices for the conformal factor $\Omega_\phi^{1-3w_\phi}(h)$ as long as it exhibits a higher power than $h^4$. 

\begin{figure*}
\centering
\includegraphics[width=0.46\textwidth]{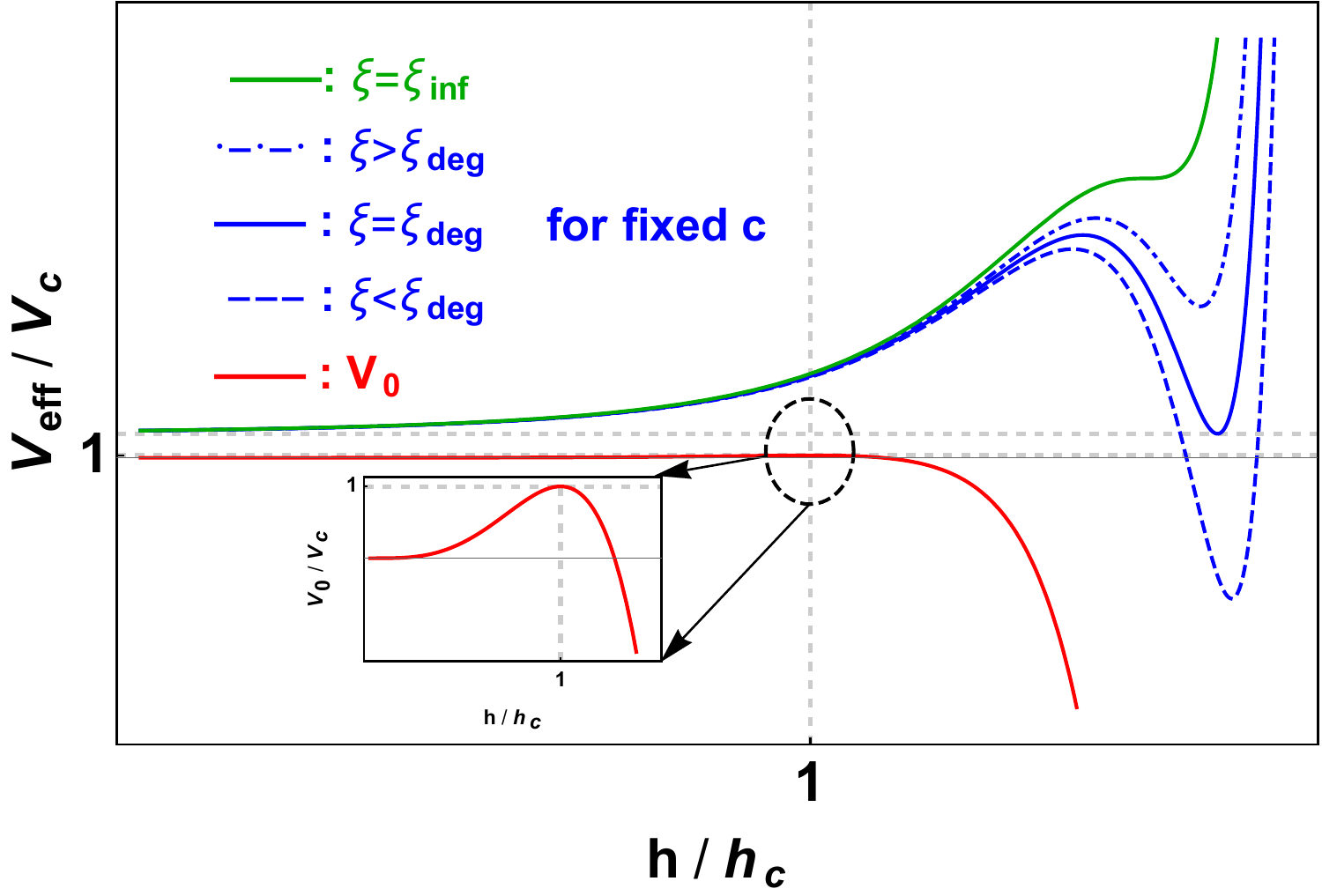}
\includegraphics[width=0.48\textwidth]{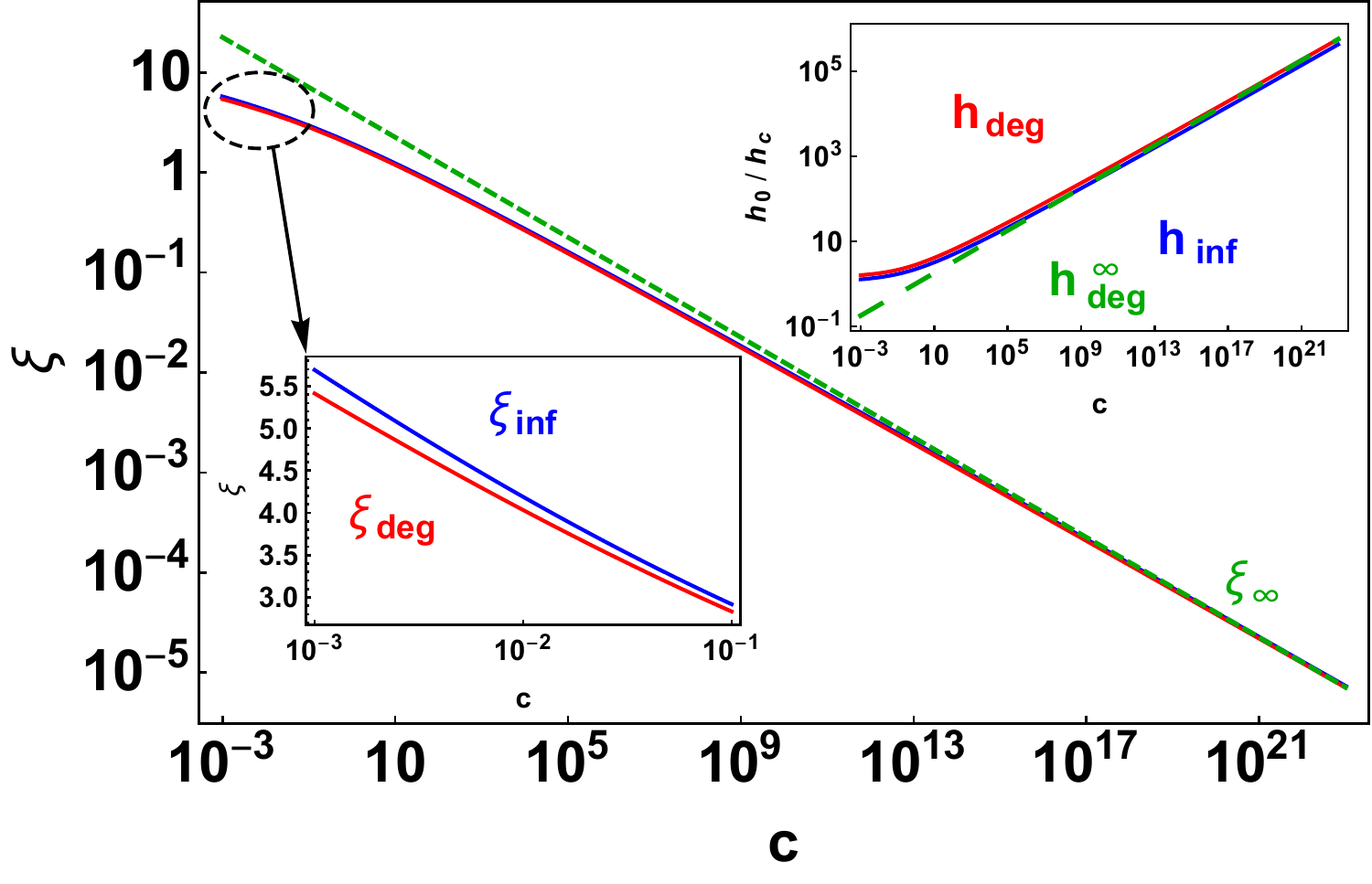}\\
\includegraphics[width=0.49\textwidth]{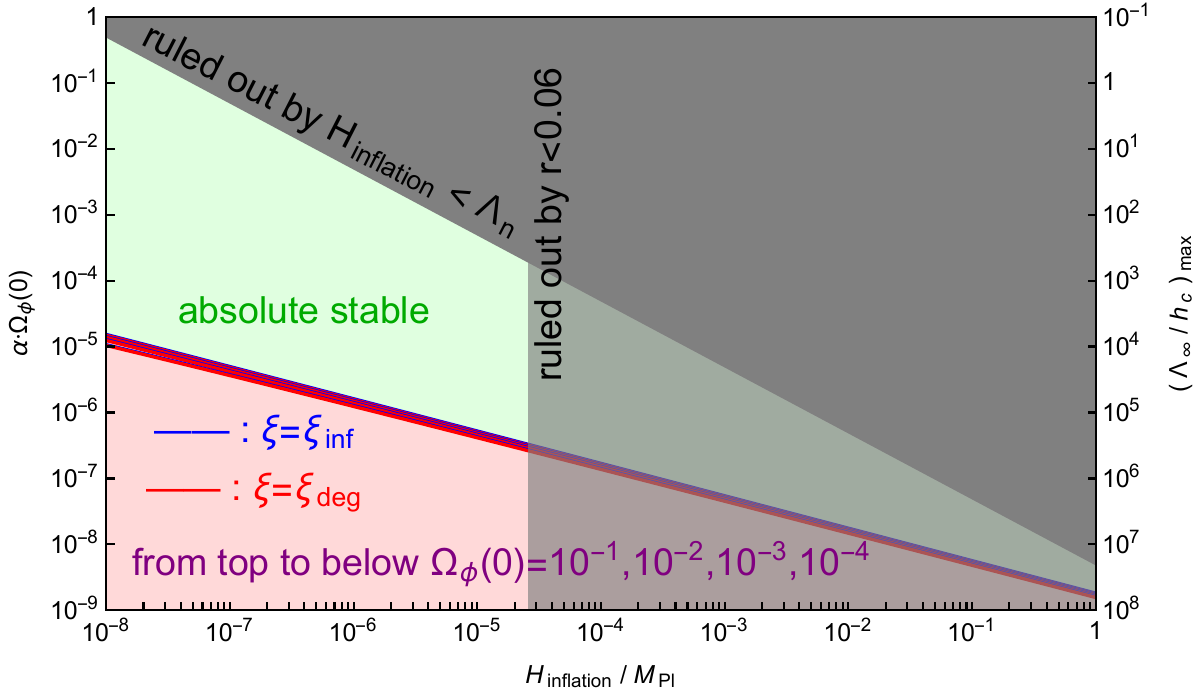}
\includegraphics[width=0.47\textwidth]{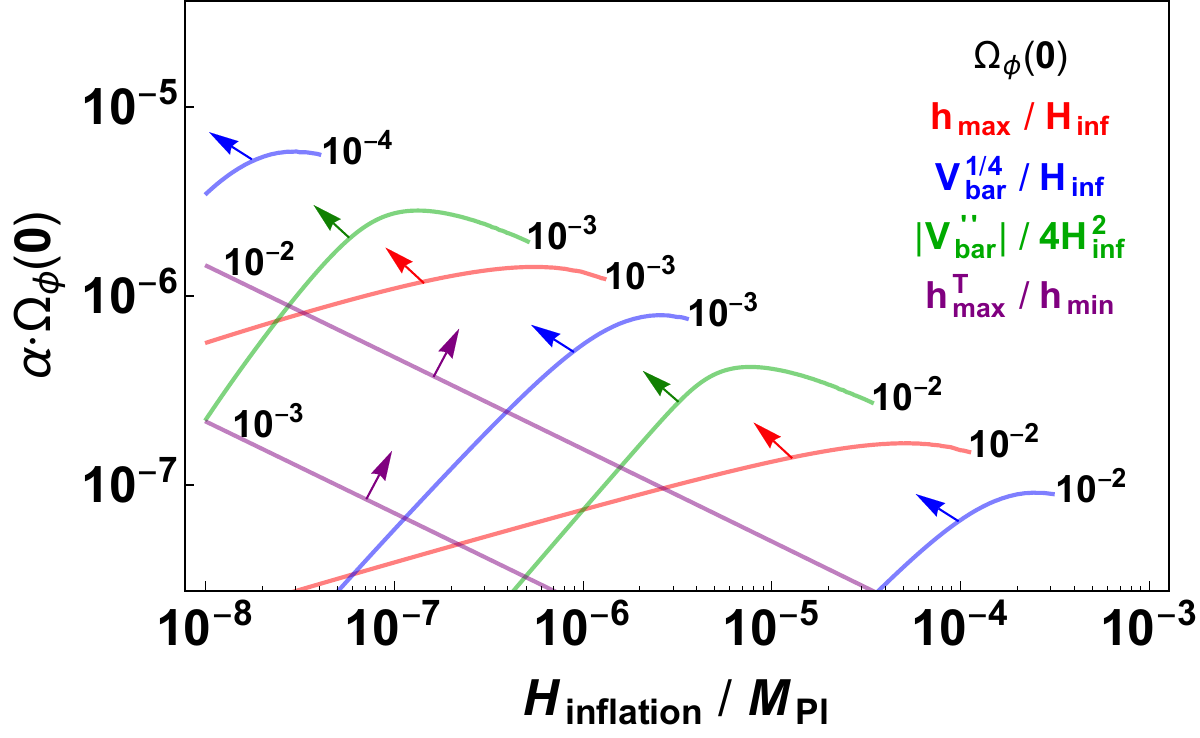}\\
\caption{\textit{Upper left :} the original unstable Higgs potential $V_0$ (red) is stabilized by the Higgs chameleon coupling to inflaton with appearance of a second minimum (blue curves around degeneracy case $\xi_\mathrm{deg}$) until its disappearance at the inflection case $\xi_\mathrm{inf}$ (green) with increasing dimensionless chameleon coupling $\alpha\equiv\xi/4$ and fixed amplitude of chameleon coupling $c$.  \textit{Upper right :} the cases of degeneracy $\xi_\mathrm{deg}$ (red) and inflection $\xi_\mathrm{inf}$ (blue) with respect to $c$ approach asymptotically to $\xi_\infty=4c^{-1/4}$ (green dashed) at large $c$ limit. The built-in panel in the lower left corner exhibits an asymptotically vanishing difference between $\xi_\mathrm{deg}$ and $\xi_\mathrm{inf}$ at large $c$ limit. The built-in panel in the upper right corner exhibits similar asymptotic behavior of Higgs field values at degenerated minimum $h_\mathrm{deg}$ (red) and inflection point $h_\mathrm{inf}$ (blue) approaches to $h_\mathrm{deg}^\infty=c^{1/4}h_c$ (green dashed) at large $c$ limit. \textit{Lower left :} the region for an absolutely stable Higgs effective potential without presence of a second minimum (green shaded) is shown above the blue lines computed from $\xi>\xi_\mathrm{inf}$ for some illustrative values of the amplitude of Higgs chameleon coupling $\Omega_\phi(0)=10^{-1}, 10^{-2}, 10^{-3}, 10^{-4}$ from top to below. The gray shaded regions are ruled out by current constraint on the tensor-to-scalar ratio $r<0.06$ and the UV effectiveness $H_\mathrm{inf}<\Lambda_n$. The stability analysis in the red shaded region below the blue lines with presence of a second minimum is presented in the next panel.  \textit{Lower right :} for given amplitude of Higgs chameleon coupling $\Omega_\phi(0)$ (black numbers), the directions of arrows point to larger position, higher height, and broader width of Higgs potential barrier with respect to Higgs quantum fluctuation scale, $h_\mathrm{max}/H_\mathrm{inf}$ (red), $V_\mathrm{bar}^{1/4}/H_\mathrm{inf}$ (blue), and $|V''_\mathrm{bar}|/(4H_\mathrm{inf}^2)$ (green) as well as  larger position of Higgs potential barrier at finite temperature with respect to the position of the second minimum at zero temperature $h_\mathrm{max}^T/h_\mathrm{min}$ (purple). }\label{fig:stabilizer}
\end{figure*}

\subsection{Dilatonic chameleon coupling}

As an illustrative example, the conformal factor could be naturally parametrized as
\begin{align}\label{eq:Omegaphi}
\Omega_\phi(h)=\Omega_\phi(0)e^{\beta h/M_\mathrm{Pl}},\quad \beta=\frac{\mathrm{d}\ln\Omega_\phi}{\mathrm{d}(h/M_\mathrm{Pl})}.
\end{align}
One could also equivalently reparametrize \eqref{eq:Omegaphi} as
\begin{align}
\Omega_\phi(h)=\Omega_\phi(0)e^{\alpha h/h_c},\quad \alpha=\frac{\mathrm{d}\ln\Omega_\phi}{\mathrm{d}(h/h_c)},
\end{align}
as long as $\alpha\equiv\beta h_c/M_\mathrm{Pl}$ is a small parameter due to hierarchy $h_c\ll M_\mathrm{Pl}$, which is indeed the case as we will see in \eqref{eq:stable1}. Note that we have implicitly assumed $h>0$ for \eqref{eq:Omegaphi}. For the region with $h<0$, one could simply allow $\beta$ to take negative value or equivalently replacing $h$ by its absolute value $|h|$ so that the rest of the paper remains unchanged.  Other even function forms (for example, quadratic in $h$ in the exponent) for the chameleon coupling are also allowed, and our specific choice only serves as an explicit illustration to manifest the mechanism.

Now the Higgs effective potential could be normalized with respect to $V_c\equiv V_0(h_c)=(b/8)h_c^4$ as
\begin{align}\label{eq:Veff}
\frac{V_\mathrm{eff}}{V_c}=-2\log\left(\frac{h^2}{h_c^2\sqrt{e}}\right)\frac{h^4}{h_c^4}+c\,e^{\xi\frac{h}{h_c}},
\end{align}
where the second term is characterized by two effective parameters defined by
\begin{align}
c\equiv\frac{\hat{\rho}_\phi}{V_c}\Omega_\phi(0)^{1-3w_\phi}, \quad \xi\equiv(1-3w_\phi)\alpha.
\end{align}
This effective potential is shown in the upper left panel of Fig. \ref{fig:stabilizer}, where the SM Higgs potential (red line) corrected by the chameleon contribution from coupling to inflaton could be easily stabilized with appearance of a second minimum (blue lines) until its disappearance at an inflection point (green line) with increasing $\xi$ or $c$. 

The second minimum $h_\mathrm{min}$ is one of the roots of the extreme points $h_0$ from $V'_\mathrm{eff}(h_0)=0$ by 
\begin{align}\label{eq:h0}
\xi\frac{h_0}{h_c}=W\left(\frac{16}{c}\frac{h_0^4}{h_c^4}\log\frac{h_0}{h_c}\right)
\end{align}
with Lambert function $W(z)$ defined by $z=W(z)e^{W(z)}$. On the one hand, for the second minimum being the degeneracy case with $V_\mathrm{eff}(h_0)=V_\mathrm{eff}(0)=cV_c$, it admits
\begin{align}\label{eq:deg}
\xi\frac{h_0}{h_c}=\frac{16(h_0/h_c)^4\log(h_0/h_c)}{c+4(h_0/h_c)^4\log(h_0/h_c)-(h_0/h_c)^4},
\end{align}
which, after combing with \eqref{eq:h0}, could solve for $\xi_\mathrm{deg}$ from given $c$ as shown in red line in the right panel of Fig. \ref{fig:stabilizer}. On the other hand, for the second minimum being the inflection point with $V''_\mathrm{eff}(h_0)=0$, it admits
\begin{align}\label{eq:inf}
\xi\frac{h_0}{h_c}=\frac{1+3\log(h_0/h_c)}{\log(h_0/h_c)},
\end{align}
which, after combing with \eqref{eq:h0}, could solve for $\xi_\mathrm{inf}$ from given $c$ as shown in blue line in the upper right panel of Fig. \ref{fig:stabilizer}. The difference between $\xi_\mathrm{deg}$ and $\xi_\mathrm{inf}$ is asymptotically vanishing at large $c$ limit, both of which are decreasing with power-law at large $c$ limit, approaching to the green dashed line, $\xi_\infty=4c^{-1/4}$, determined by first solving $\log(h_\mathrm{deg}/h_c)$ as a whole from \eqref{eq:deg} and then plugging into \eqref{eq:h0} with asymptotic expansion of Lambert function $W(z\to0)\sim z+\mathcal{O}(z^2)$. The corresponding $h_\mathrm{deg}/h_c$ in the $c\to\infty$ limit approaches $c^{1/4}$. 

\subsection{Absolutely stable region}

Without the appearance of the second minimum when $\xi>\xi_\mathrm{inf}$, the Higgs field is absolutely stable against any quantum fluctuations. For large enough $c$, the absolutely stable region could be approximately estimated by
\begin{align}\label{eq:stable0}
\xi>\xi_\mathrm{inf}\approx\xi_\mathrm{deg}\sim\xi_\infty=4c^{-1/4}.
\end{align}
To further transform the above constraints on $(c,\xi)$ into more physical constraints on the inflationary Hubble scale $H_\mathrm{inf}$ and the dimensionless conformal factor $\alpha$, we could first set the EoS parameter $w_\phi=-1$ during inflation without loss of generality, then $\alpha=\xi/4$ and $c$ is related to $H_\mathrm{inf}$ by
\begin{align}\label{eq:c}
c=\frac{3M_\mathrm{Pl}^2H_\mathrm{inf}^2}{V_c}\Omega^4_\phi(0)=\frac{24}{b}\left(\frac{M_\mathrm{Pl}}{h_c}\right)^4\left(\frac{H_\mathrm{inf}}{M_\mathrm{Pl}}\right)^2\Omega_\phi^4(0).
\end{align}
To ensure that the Higgs effective potential energy $V_\mathrm{eff}(0)/V_c\equiv c$ at the desirable stable vacuum $h=0$ is sub-dominated to the background Hubble expansion, namely $c\ll3M_\mathrm{Pl}^2H_\mathrm{inf}^2/V_c$, the amplitude of conformal factor should be small, $\Omega_\phi(0)\ll1$ . Now the absolute stability condition $\xi\gtrsim\xi_\infty$ reads
\begin{align}\label{eq:stable1}
\frac{\alpha\Omega_\phi(0)}{1.6\times10^{-7}}\gtrsim\left(\frac{b}{10^{-3}}\right)^\frac14\left(\frac{h_c}{10^{10}\,\mathrm{GeV}}\right)\left(\frac{H_\mathrm{inf}}{10^{13}\,\mathrm{GeV}}\right)^{-\frac12}.
\end{align}
This suggests an absolute stability bound by the product $\alpha\cdot\Omega_\phi(0)$ in power law with respect to the inflationary Hubble scale shown as the green region in the lower left panel of Fig. \ref{fig:stabilizer}, which, without adopting the asymptotic form $\xi_\infty=4c^{-1/4}$, is precisely computed by $\xi>\xi_\mathrm{inf}$ with respect to the inflection case (blue lines) for $\Omega_\phi(0)=10^{-1}, 10^{-2}, 10^{-3}, 10^{-4}$ from top to below. Nevertheless, for given $\Omega_\phi(0)$, the corresponding red shaded region below $\xi=\xi_\mathrm{inf}$ is NOT everywhere unstable as specified below.

\subsection{UV effectiveness}

To check the UV effectiveness of our Higgs chameleon mechanism, we first expand the dilatonic coupling term as
\begin{align}
V_\mathrm{eff}=V_	0(h)+\sum\limits_n\frac{1}{n!}\left(\frac{h}{\Lambda_n}\right)^{n-4}h^4,
\end{align}
where the cutoff scale
\begin{align}
\Lambda_n\equiv\left(\frac{8}{bc\xi^n}\right)^\frac{1}{n-4}h_c,
\end{align}
after using $\xi=4\alpha=4\beta h_c/M_\mathrm{Pl}$ and replacing $c$ with \eqref{eq:c}, becomes
\begin{align}
\Lambda_n=\left(\frac13\right)^\frac{1}{n-4}\left(\frac{1}{4\beta}\right)^\frac{n}{n-4}\left(\frac{H_\mathrm{inf}}{M_\mathrm{Pl}}\right)^{-\frac{2}{n-4}}\Omega_\phi(0)^{-\frac{4}{n-4}}M_\mathrm{Pl}.
\end{align}
Further appreciating the absolute stability condition $\xi\gtrsim\xi_\infty$ [see \eqref{eq:stable1}]  in terms of $\beta$, namely,
\begin{align}
\beta\gtrsim\left(\frac{b}{24}\right)^\frac14\left(\frac{H_\mathrm{inf}}{M_\mathrm{Pl}}\right)^{-\frac12}\Omega_\phi^{-1}(0),
\end{align}
the cutoff scale for nonrenormalizable operators $(n>4)$ is close to the Planck scale suppressed by the parameter $\beta$,
\begin{align}
\Lambda_n\lesssim\left(\frac{8}{4^nb}\right)^\frac{1}{n-4}\frac{M_\mathrm{Pl}}{\beta},
\end{align}
where the prefactor $[8/(4^nb)]^{1/(n-4)}$ approaches $1/4$ from above in the large $n$ limit. Since our Higgs chameleon mechanism is proposed to address the Higgs metastability problem, the lowest cutoff scale $\Lambda_{n\to\infty}=M_\mathrm{Pl}/(4\beta)$ should at least larger than the Higgs instability scale $h_c$. We therefore label the maximum value of $\Lambda_\infty/h_c<1/(4\alpha)$ in the lower left panel of Fig. \ref{fig:stabilizer} for given $\alpha\Omega_\phi(0)$ with $\Omega_\phi(0)<1$. It is easy to see in the green shaded region that the cutoff scale is not that far above the Higgs instability scale. 
On the other hand, to ensure the effectiveness of our scenario during inflation, one should also impose the condition $H_\mathrm{inf}<\Lambda_n$ that the cutoff scale should be larger than the characteristic inflationary scale, namely,
\begin{align}
\beta<\left(\frac{8}{4^nb}\right)^\frac{1}{n-4}\frac{M_\mathrm{Pl}}{H_\mathrm{inf}}.
\end{align}
Since $[8/(4^nb)]^{1/(n-4)}$  is always larger than $1/4$ and $M_\mathrm{Pl}\gg H_\mathrm{inf}$, this condition could be easily fulfilled. If this condition should be satisfied for all $n$, then one only needs to require
\begin{align}
\beta<\frac14\frac{M_\mathrm{Pl}}{H_\mathrm{inf}}\Leftrightarrow\alpha<\frac14\frac{h_c}{H_\mathrm{inf}}.
\end{align}
Since the background expansion is dominated by the inflaton field by $\Omega_\phi(0)\ll1$, this also puts an upper bound on $\alpha\Omega_\phi(0)$ shown as the gray shaded region in the third panel of Fig. \ref{fig:stabilizer}. As an illustrative benchmark example, one could take $\alpha\Omega_\phi(0)\sim10^{-5}$ inside the absolute stability regime for $H_\mathrm{inf}/M_\mathrm{Pl}\sim10^{-6}$, thus $\alpha<5\times10^{-3}$, and one only needs to choose $\Omega_\phi(0)\gtrsim2\times10^{-3}$. 

\subsection{Presence of a second minimum}

The second minimum appears when $\xi<\xi_\mathrm{inf}$, which is higher or lower than the $h=0$ vacuum if $\xi_\mathrm{deg}<\xi<\xi_\mathrm{inf}$ or $\xi<\xi_\mathrm{deg}$, respectively. The degeneracy cases $\xi=\xi_\mathrm{deg}$ are shown as red lines in the lower left panel of Fig. \ref{fig:stabilizer} for $\Omega_\phi(0)=10^{-1}, 10^{-2}, 10^{-3}, 10^{-4}$ from top to below.  In the presence of a second minimum, the Higgs stability  against quantum fluctuations is guaranteed in all $e^{3N_0}$ Hubble patches in our past light cone if \cite{Espinosa:2015qea,Kohri:2016wof}
\begin{align}\label{eq:stable2}
\frac{h_\mathrm{max}}{H_\mathrm{inf}}>n_\mathrm{stab}\equiv\begin{cases}
\frac{3\sqrt{N_0}}{2\pi}\frac{H_\mathrm{inf}}{m_\mathrm{eff}}, & m_\mathrm{eff}<\frac32H_\mathrm{inf},\\
\sqrt{\frac{N_0}{2\pi^2}\frac{H_\mathrm{inf}}{m_\mathrm{eff}}}, & m_\mathrm{eff}>\frac32H_\mathrm{inf},
\end{cases}
\end{align}
where $h_\mathrm{max}$ is the other root of \eqref{eq:h0}, $N_0\approx60$ is the e-folding number of our current Hubble scale leaving the Hubble horizon before the end of inflation, and $m_\mathrm{eff}$ is given by
\begin{align}\label{eq:meff0}
m_\mathrm{eff}^2(h=0)\equiv V''_\mathrm{eff}(h=0)=\frac{bc\xi^2}{8}h_c^2.
\end{align}
For given $\Omega_\phi(0)=10^{-2}, 10^{-3}, 10^{-4}$ (black numbers) in the lower right panel of Fig. \ref{fig:stabilizer}, we have tested the condition \eqref{eq:stable2} as red curves with red arrows pointing to a larger value than $n_\mathrm{stab}$, which automatically guarantees a much higher potential barrier $V_\mathrm{bar}\equiv V_\mathrm{eff}(h_\mathrm{max})-V_\mathrm{eff}(0)>H_\mathrm{inf}^4$ (blue curves) than the inflationary Hubble scale for the same $\Omega_\phi(0)$. This largely suppresses the decay processes via either CdL instanton or Hawking-Moss (HM) instanton depending on the broadness of potential barrier estimated by $|V''_\mathrm{eff}(h_\mathrm{max})|/(4H_\mathrm{inf}^2)$ \cite{Hook:2014uia} (green curves), to the upper-left/lower-right of which are dominated by CdL/HM instantons (if ever happened via decay channel), respectively. Therefore, the Higgs stability region against the quantum fluctuations could be extended from the absolutely stable region (green shaded) into the red shaded region in the lower left panel of Fig. \ref{fig:stabilizer} bounded by the red curves in the lower right panel of Fig. \ref{fig:stabilizer} for given $\Omega_\phi(0)$.

However, this is not the whole story. Even for the parameter region to the lower-right direction of red curve with given $\Omega_\phi(0)$ where the second minimum is accidentally achieved during inflation either by the rare decay instantons or random walks over the potential barrier in some of the Hubble patches, there is still hope for them to be saved by the thermal corrections to the Higgs potential during radiation dominated era as elaborated below.

\subsection{Thermal rescue}

For an instantaneous reheating history, the reheating temperature at the onset of radiation domination approximately reads from the inflationary energy, 
\begin{align}
\frac{T_\mathrm{reh}}{M_\mathrm{Pl}}\approx\left(\frac{90}{g_\mathrm{reh}\pi^2}\right)^{1/4}\left(\frac{H_\mathrm{inf}}{M_\mathrm{Pl}}\right)^{1/2},
\end{align}
with the number of degrees of freedom $g_\mathrm{reh}=106.75$ for SM. The Higgs effective potential simply reads $V_\mathrm{eff}(h)=V_0(h)+V_T(h)+\hat{\rho}_r$ with $\hat{\rho}_r$ independent of $h$ ($\hat{\rho}_r$ could be chosen as zero since the trace of energy-momentum tensor in \eqref{eq:HiggsEoM} is vanished for radiation dominance),  and the thermal corrections could be conveniently approximated up to $h\lesssim 2\pi T$ by $V_T(h)\approx\frac12M_T^2h^2$ with  \cite{Espinosa:2015qea}
\begin{align}
M_T^2\approx\left(0.21-0.0071\lg\frac{T}{\text{GeV}}\right)T^2,
\end{align}
which pushes the potential barrier to larger position,
\begin{align}
h_\mathrm{max}^T=M_T\left[bW\left(\frac{M_T^2}{bh_c^2}\right)\right]^{-1/2}.
\end{align} 
The thermal rescue  \cite{Espinosa:2015qea} occurs when the local maximum $h_\mathrm{max}^T$ at finite temperature $T_\mathrm{reh}$ is large enough for the Higgs field in the second minimum $h_\mathrm{min}$ achieved during inflation could subsequently roll back to $h=0$ vacuum during radiation era,
\begin{align}
h_\mathrm{max}^T(T_\mathrm{reh})>h_\mathrm{min},
\end{align}
which is shown as purple curves in the lower right panel of Fig. \ref{fig:stabilizer} with the direction of arrows pointing to the larger ratio of $h_\mathrm{max}^T/h_\mathrm{min}$ than unity value. After the thermal rescue, the thermal fluctuations of order temperature $T$ have been checked to be much smaller than the thermal potential barrier, $h_\mathrm{max}^T\gg T$.

For noninstantaneous reheating, $U_i(h)$ in \eqref{eq:Uh} during pre/reheating is smaller than that from inflationary era due to smaller power $1-3w_i<4$ with $-1/3<w_i<1/3$ and smaller $\hat{\rho}_i$ that dissipates into radiations, which could push the second minimum (if ever reached during inflation) to larger and deeper values until gradually connecting to the thermal Higgs potential in radiation era, thus invalidating the thermal rescue mechanism. Furthermore, one still has to avoid the broad resonance even though the positive effective mass-squared at either $h=0$ vacuum or the second minimum could evade the tachyonic resonant production of Higgs during preheating. Therefore, a conservative safe zone is that $V_\mathrm{eff}(h)$ never develops a second minimum to be ever reached during inflation and relaxed during pre/reheating, namely \eqref{eq:stable1}. We hope to revisit this issue in more details in a separate paper in future.

\section{Conclusion and discussions}\label{sec:con}

We have proposed a new mechanism to stabilize the Higgs potential in the early Universe by regarding Higgs as chameleon coupled to inflaton, which simply adds positive contribution to the original Higgs potential as shown in \eqref{eq:HiggsChameleon}. We have tested this proposal in an illuminating example with conformal factor of form exponential to Higgs field as shown in \eqref{eq:Veff}. Other forms of this conformal factor should also work as long as it contributes positively to the effective potential.  The absolutely stability bound \eqref{eq:stable0}, or expressed in terms of inflationary Hubble scale as \eqref{eq:stable1}, is analytically derived from the disappearance of inflection point in the effective potential. We also preliminarily extended the stability regime beyond the absolutely stable region into the case with the presence of a second minimum. Several comments are in order below.

First, our solution for the Higgs stability problem in the early Universe only requires a chameleon coupling of Higgs to inflaton alone, while the chameleon couplings of Higgs to other fields are not necessarily demanded, which buys us extra benefit of evading all the current constraints on Higgs from either particle colliers or local gravity experiments. 

Second, our identification of Higgs boson as chameleon field serves as a phenomenological model, whose ultraviolet (UV) completion goes beyond the scope of current goal for resolving SM metastability issue. Nevertheless, a UV completion \cite{Hinterbichler:2010wu} of general chameleon could be realized by identifying chameleon scalar field with a certain function of the volume modulus of the extra dimensions. Therefore, embedding Higgs in extra dimensions \cite{Hosotani:1983xw} is a promising starting point for the UV completion of Higgs chameleon.

Third, we neglect the effects on the running of SM Higgs couplings from Higgs-inflaton chameleon-like coupling, which, after expanding the conformal factor in power of $h$, only contributes to SM Higgs couplings with terms proportional to the same power of product $\alpha\Omega_\phi(0)$, which is quite small ($\delta m^2\sim10^{-14}, \delta\lambda\sim10^{-28}$) according to the typical value of the absolute stability bound \eqref{eq:stable1}.

Finally, three possible traces of Higgs ever as chameleon in the early Universe could be the isocurvature perturbations and non-Gaussianity due to its chameleon coupling to inflaton, as well as the productions of domain walls \cite{Deng:2016vzb,Liu:2019lul,Deng:2017uwc} (see also \cite{Kusenko:2020pcg}) when the second minimum is accidentally achieved during inflation in some Hubble patches, which merits further studies in the future.

\begin{acknowledgments}
We thank Mark Hertzberg, Justin Khoury, Jing Liu, Shan-Ming Ruan, Zhong-Zhi Xianyu,  Run-Qiu Yang and Yue Zhao for helpful correspondences. We also thank an anonymous referee for raising the issue of the UV effectiveness.
R.G.C. was supported by the National Natural Science Foundation of China Grants No. 11947302, No. 11991052, No. 11690022, No. 11821505 and No. 11851302, and by the Strategic Priority Research Program of CAS Grant No. XDB23030100, and by the Key Research Program of Frontier Sciences of CAS. S.J.W. is supported by the postdoctoral scholarship of Tufts University from NSF.
\end{acknowledgments}


\bibliographystyle{utphys}
\bibliography{ref}

\end{document}